\documentclass[12pt]{article}
\pdfoutput=1

\usepackage{putex}
\usepackage{graphicx}
\usepackage{caption}
\usepackage{amsmath}
\usepackage{array}
\usepackage{subcaption}
\usepackage{epstopdf}
\usepackage{enumerate}
\usepackage{cite}
\usepackage{youngtab}
\usepackage{tensor}
\usepackage{slashed}
\usepackage[utf8]{inputenc}
\usepackage{rotating}
\usepackage{bigfoot}
\usepackage[
      colorlinks=true,
      linkcolor=blue,
      urlcolor=blue,
      filecolor=black,
      citecolor=red,
      ]{hyperref}

\newcommand\mc[1]{\mathcal{#1}}

\newcommand {\be} {\begin {equation}}
\newcommand {\ee} {\end {equation}}

\newcommand {\bes} {\begin {equation*}}
\newcommand {\ees} {\end {equation*}}

\newcommand{\es}[2] {\begin{equation} \label{#1} \begin{split} #2 \end{split} \end{equation}}

\newcommand{\cA}{{\mathcal A}}

\newcommand{\cO}{{\mathcal O}}

\newcommand{\beq}{\begin{equation}}
\newcommand{\eeq}{\end{equation}}

\def\ie{\begin{equation}\begin{aligned}}
\def\fe{\end{aligned}\end{equation}}

\newcommand{\mC}{{\mathbb C}}

\newcommand{\mZ}{{\mathbb Z}}

\numberwithin{equation}{section}

\def\<{\langle}
\def\>{\rangle}

\begin{document}

\preprint{}

\institution{weizmann}{Department of Particle Physics and Astrophysics, Weizmann Institute of Science, Rehovot, Israel}

\title{3d $\mathcal{N}=4$ OPE Coefficients from Fermi Gas}

\authors{Shai M. Chester\worksat{\weizmann}, Rohit R. Kalloor\worksat{\weizmann}, and Adar Sharon\worksat{\weizmann}}

\abstract{
The partition function of a 3d $\mathcal{N}=4$ gauge theory with rank $N$ can be computed using supersymmetric localization in terms of a matrix model, which often can be formulated as an ideal Fermi gas with a non-trivial one-particle Hamiltonian. We show how OPE coefficients of protected operators correspond in this formalism to averages of $n$-body operators in the Fermi gas, which can be computed to all orders in $1/N$ using the WKB expansion. We use this formalism to compute OPE coefficients in the $U(N)_k\times U(N)_{-k}$ ABJM theory as well as the $U(N)$ theory with one adjoint and $N_f$ fundamental hypermultiplets, both of which have weakly coupled M-theory duals in the large $N$ and finite $k$ or $N_f$ regimes. For ABJM we reproduce known results, while for the $N_f$ theory we compute the all orders in $1/N$ dependence at finite $N_f$ for the coefficient $c_T$ of the stress tensor two-point function.
}
\date{}

\maketitle

\tableofcontents

\section{Introduction}
\label{intro}

Supersymmetric localization is one of the few available tools for analytically computing nontrivial physical observables non-perturbatively in interacting quantum field theories. In the original case of 4d $\mathcal{N}=4$ super-Yang-Mills (SYM) with gauge group $SU(N)$, Pestun used localization to compute the sphere partition function and the expectation value of Wilson loops in terms of $(N-1)$-dimensional matrix model integrals \cite{Pestun:2007rz}. Localization was then applied by \cite{Kapustin:2009kz} to 3d $\mathcal{N}=2$ Chern-Simons-matter theories, such as the $\mathcal{N}=6$ ABJM theory with gauge group $U(N)_k\times U(N)_{-k}$  and Chern-Simons level $k$ \cite{Aharony:2008ug}, to compute matrix model observables as $N$-dimensional integrals. For low $N$, these integrals can be computed by hand, but this becomes infeasible for large $N$. The large $N$ regime is of particular interest for conformal field theories (CFTs) with holographic duals, such as 4d $\mathcal{N}=4$ SYM dual to Type IIB string theory on $AdS_5\times S^5$, and 3d ABJM dual to Type IIA string theory on $AdS_4\times\mathbb{CP}^3$ for large $k$ and M-theory on $AdS_4\times S^7/\mathbb{Z}_k$ at small $k$. When $N$ is large, the bulk is described by weakly coupled supergravity, so if the localization integrals can be computed at large $N$, then they could be compared to bulk calculations. For $\mathcal{N}=4$ SYM, the matrix model is Gaussian, so observables in the matrix model can be computed to all orders in $1/N$ or even finite $N$ using standard matrix model methods like orthogonal polynomials and topological recursion and then successfully matched to bulk calculations \cite{Drukker:2000rr,Fiol:2013hna,Chester:2019pvm,Chester:2020dja,Chester:2019jas}. For ABJM  the matrix model is much more complicated, but nevertheless topological recursion was successfully applied in \cite{Drukker:2010nc} to compute matrix model observables in the 't Hooft limit of large $N,k$ and fixed $N/k$. This limit could not be used to study the M-theory regime, however, which is only weakly coupled in the bulk for large $N$ and finite $k$.

To study the M-theory limit of ABJM, \cite{Marino:2011eh} showed that the ABJM matrix model could be written in the form
\es{Zintro}{
Z=\frac{1}{N!}\sum_{\sigma\in S_N}(-1)^{|\sigma|}\int d^N x\langle x_1\dots x_N|\bigotimes_{i=1}^N\hat\rho|x_{\sigma(1)}\dots x_{\sigma(N)}\rangle\,,
}
which describes a gas of non-interacting fermions with a certain 1-particle density matrix $\rho$ and $\hbar\sim k$. The partition function could then be computed in a small $\hbar\sim k$ WKB expansion using the Wigner approach to quantum statistical mechanics, and the all orders in $1/N$ result could be written in terms of an Airy function. At leading order in large $N$, the resulting sphere free energy scaled as $N^{\frac32}$ as predicted from M-theory. Wilson loops could also be computed as thermal averages of 1-body operators in the Fermi gas \cite{Klemm:2012ii}, and the all orders in $1/N$ result also took the form of Airy functions. This Fermi gas method was soon generalized to a wide variety of 3d $\mathcal{N}=4$ theories (for a review with extensive references, see \cite{Marino:2016new}). For instance, the $\mathcal{N}=4$ $U(N)$ gauge theory coupled to one hypermultiplet in the adjoint representation and $N_f$ hypermultiplets in the fundamental is holographically dual to M-theory on $AdS_4\times S^7/\mathbb{Z}_{N_f}$ for finite $N_f$ \cite{Bashkirov:2010kz,Benini:2009qs}. In \cite{Grassi:2014vwa,Hatsuda:2014vsa}, the sphere partition function of this $N_f$ matrix model was written in the form of \eqref{Zintro}, and was then computed to all orders in $1/N$ in terms of Airy functions. 

Applications of the Fermi gas method have so far been limited to the sphere partition function and the expectation value of Wilson loops. Recently, it has become clear that a much wider class of physical observables can be computed using localization. One such class are integrated correlators of conserved currents. Since localization does not require conformal symmetry, the sphere partition function can be computed for a theory deformed by real masses $m$, which couple to flavor multiplets, as well as on the squashed sphere with squashing parameter $b$ \cite{Hama:2010av,Hama:2011ea,Imamura:2011wg}, which couples to the stress tensor multiplet. Taking $n$ derivatives of the free energy in terms of $m$ (or $b$) and then setting $m=0$ (or $b=1$) then computes the correlator of $n$ flavor currents (or stress tensors) integrated over the sphere. For instance, consider the coefficient $c_T$ of the two-point function of canonically normalized stress-tensors:
 \es{CanStress}{
  \langle T_{\mu\nu}(\vec{x}) T_{\rho \sigma}(0) \rangle = \frac{c_T}{64} \left(P_{\mu\rho} P_{\nu \sigma} + P_{\nu \rho} P_{\mu \sigma} - P_{\mu\nu} P_{\rho\sigma} \right) \frac{1}{16 \pi^2 \vec{x}^2} \,, \qquad P_{\mu\nu} \equiv \eta_{\mu\nu} \nabla^2 - \partial_\mu \partial_\nu \,,
 }
 which is related by the conformal Ward identity to the OPE coefficient $\lambda_{\cO\cO T}$  for any scalar operator $\cO$ as $c_T\propto 1/\lambda_{\cO\cO T}^2$ \cite{Osborn:1993cr}. We can compute $c_T$ by taking two derivatives of $b$ as
   \es{cTGen}{
c_T=&-\frac{32}{\pi^2}{\partial^2_{b}\log Z}\big\vert_{b=1}\,,
 }
 which expresses $c_T$ in terms of the expectation value of a certain 2-body operator in the matrix model. This method was used in \cite{Closset:2012ru,Nishioka:2013gza,Nishioka:2013haa} to compute $c_T$ in a wide variety of 3d $\mathcal{N}=2$ theories whose matrix models involved only a small number of integrals that could be performed by hand. For ABJM, the three real masses $m_i$ couple to the R-symmetry currents that are in the same multiplet as the stress tensor, so $c_T$ can be computed from $\partial_{m_i}^2\log Z\big\vert_{m=0}$. The mass deformed partition function $Z(m_i)$ was computed to all orders in $1/N$ using the Fermi gas method in \cite{Nosaka:2015iiw}, which was used in \cite{Agmon:2017xes} to compute $c_T$ to all orders in $1/N$. The integrated stress tensor multiplet 4-point function was similarly computed from $\partial_{m_i}^2\partial_{m_j}^2\log Z\big\vert_{m=0}$ in \cite{Binder:2018yvd} and used to constrain the correlator in the large $N$ expansion.

Another class of observables that can be computed using matrix model expectation values are OPE coefficients of half-BPS scalar operators in 3d $\mathcal{N}=4$ CFTs. As shown in \cite{Chester:2014mea,Beem:2016cbd}, all 3d $\mathcal{N}=4$ CFTs contain a 1d topological sector that describes half-BPS operators in the theory. For certain theories, such as ABJM with $k=1$, a Lagrangian was derived for this 1d sector that could be used to compute OPE coefficients of half-BPS operators as expectation values of $n$-body operators in the matrix model \cite{Dedushenko:2016jxl}. For small $N$, the integrals in these expectation values were evaluated explicitly in \cite{Agmon:2017lga,Agmon:2019imm}. A certain half-BPS operator that appears in the stress-tensor correlator was related to $\partial_{m_i}^4\log Z\big\vert_{m=0}$ using the 1d theory in \cite{Agmon:2017xes}, so this OPE coefficient could be computed to all orders in $1/N$ for any $k$ using the Fermi gas result for $Z(m_i)$.

In this paper, we will show how to use the Fermi gas method to compute these $n$-body operator expectation values to all orders in $1/N$ in terms of Airy functions. We will first apply these methods to  $\partial_{m_i}^2\log Z\big\vert_{m=0}$ and $\partial_{m_i}^4\log Z\big\vert_{m=0}$ for ABJM theory, where we consider these quantities as expectation values of $n$-body operators with respect to the $m_i=0$ matrix model. Our results match those previously computed using the Fermi gas expression for $Z(m_i)$, which serves as a non-trivial check of our methods. We will then consider $c_T$ as computed from ${\partial^2_{b}\log Z}\big\vert_{b=1}$ in the $N_f$ matrix model described above. We will compute the all orders in $1/N$ result for this quantity up to an additive $N$-independent constant, which in particular allows us to extract the first two leading orders at large $N$ and finite $N_f$. This is a new result that could be compared to bulk calculations in the future.

  The rest of this paper is organized as follows.  In Section~\ref{nbody}, we discuss the calculation of $n$-body operator expectation values using the Fermi gas technique, and explain why the answers take the form of Airy functions. In Section~\ref{abjm} we then apply these methods to ABJM theory and recover the known results for $\partial_{m_i}^2\log Z\big\vert_{m=0}$ and $\partial_{m_i}^4\log Z\big\vert_{m=0}$. In Section~\ref{nf} we apply our methods to the $N_f$ matrix model to compute $c_T$.  We end with a discussion of our results and future directions in Section~\ref{conc}.  Details of our calculations are given in various Appendices.

\section{Fermi gas for $n$-body operators}
\label{nbody}

We begin by reviewing the calculation of quantum-mechanical averages of $n$-body operators in a free Fermi gas, using the Wigner formalism. This is a classic subject that has been recently reviewed in \cite{Klemm:2012ii}, so we will be brief. For the $n$-body operators that we consider for 3d $\mathcal{N}=4$ matrix models, we show that these averages in the grand canonical ensemble are polynomial in the chemical potential $\mu$, so that the $N$-dependence of the canonical ensemble averages are captured by Airy functions, as was previously shown for the partition function.

Let us first consider a system of $N$ distinguishable particles, whose Hilbert space is spanned by the basis of eigenstates $|x_1,\dots,x_N\rangle$. An $n$-body operator $\cO_n$ is defined as an operator that is invariant under any permutation of the particles and acts on the Hilbert space as
\es{ndbodyO}{
\cO_n |x_1,\dots,x_N\rangle=\frac{1}{n!}\sum_{1\leq i_1\neq\cdots\neq i_n\leq N}\cO(x_{i_1},\dots, x_{i_n})|x_1,\dots,x_N\rangle\,.
} 
For a system of $N$ identical fermions, we must anti-symmetrize the states using the projection operator
\es{P}{
P_N=\frac{1}{N!}\sum_{\sigma\in S_N}(-1)^{|\sigma|}\sigma\,.
}

As discussed in the Introduction, we want to compute the thermal averages of these operators in the canonical ensemble of a free Fermi gas with  temperature $T=1$. The canonical density matrix $\hat \rho^\text{gas}$ in a free gas factorizes into single particle density matrices $\hat\rho$ as
\es{rhotot0}{
\hat \rho^\text{gas}=\bigotimes_{i=1}^N\hat\rho\,,\qquad\qquad \hat\rho=e^{-\hat H}\,,
}
where $\hat H$ is the 1-particle Hamiltonian. For the thermal average of an $n$-body operator, we can integrate out $(N-n)$ particles to write the average in terms of the $n$-particle density matrix $\rho_n$ as
\es{thermalAv}{
\langle \cO_n\rangle&=\frac{1}{n!}\int dx_1\cdots dx_n \cO(x_1,\dots,x_n)\rho_n(x_1,\dots,x_n)\,,\\
\rho_n(x_1,\dots,x_n)&\equiv\frac{N!}{(N-n)!}\int dx_{n+1}\cdots dx_N  \langle x_1,\dots,x_N|P_N\bigotimes_{i=1}^N\hat\rho|x_1,\dots,x_N\rangle\,,
}
where $\cO_n$ is now restricted to the $n$-particle Hilbert space. These thermal averages are unnormalized, so that the canonical partition function $Z_N$ in \eqref{CanStress} is by definition the average of the zero-body unit operator $Z_N=\langle 1\rangle$.

The canonical ensemble is difficult to use in practice due to the restriction to a definite number $N$ of particles. Instead, we can eliminate $N$ using a Legendre transform to define the grand canonical ensemble with chemical potential $\mu$, so that averages in each ensemble are related as
\es{grand}{
\langle\cO_n\rangle^\text{GC}(\mu)=\sum_{N=n}^\infty \langle\cO_n\rangle(N) e^{\mu N}\,,\qquad\Leftrightarrow\qquad \langle \cO_n\rangle(N)=\frac{1}{2\pi i}\int d\mu e^{-\mu N}\langle\cO_n\rangle^\text{GC}(\mu)\,.
}
For instance, the grand partition function $\Xi$ corresponds to the average of the zero-body unit operator, and so takes the form
\es{xi}{
\Xi(\mu)=1+\sum_{N=1}^\infty Z(N) e^{\mu N}\,,\qquad\Leftrightarrow\qquad  Z(N)=\frac{1}{2\pi i}\int d\mu e^{-\mu N}\Xi(\mu)\,.
}
As in the canonical ensemble, the density matrix of a free Fermi gas factorizes and can be written in terms of single particle Fermi distributions as
\es{rhotot}{
\hat \rho^\text{gas,GC}=\Xi\bigotimes_{i=1}^N\hat\rho_\text{GC}(\mu)\,,\qquad\qquad \hat\rho_\text{GC}(\mu)=\frac{1}{e^{(\hat H-\mu)}+1}\,.
}
 The grand canonical average of an $n$-body operator in a free Fermi gas can then be written in the $n$-particle Hilbert space as
\es{traceGC}{
\langle \cO_n \rangle^\text{GC}(\mu)=\Xi\int d^nx \langle x_1,\dots,x_n|P_n\cO_n\bigotimes_{i=1}^n \hat\rho_\text{GC}(\mu)|x_1,\dots,x_n\rangle\,.
}

This average can be computed in a semiclassical small $\hbar$ expansion using the Wigner formalism. The Wigner transform of an $n$-body operator $\cO_n$ is defined as
\es{antiW}{
(\cO_n)_W=\prod_{i=1}^n\int dy_i\langle x_i-\frac{y_i}{2}|\cO_n|x_{i}+\frac{y_{i}}{2}\rangle e^{\frac{i p_i y_i}{\hbar}}\,,
}
where $p_i$ is the canonical conjugate of $x_i$ with the standard commutation relations
\es{hbar}{
[x_i,p_j]=i\hbar\delta_{ij}\,.
}
The Wigner transform satisfies the product rule
\es{wigProd}{
\left(\hat A\hat B\right)_W=A_W\star B_W\,,\qquad \star\equiv\exp\left[\frac{i\hbar}{2}\left(\overleftarrow{\partial_x}\overrightarrow{\partial _p}-\overleftarrow{\partial_p}\overrightarrow{\partial_x}\right)\right]\,.
}
The utility of the Wigner transform is that it computes quantum statistical averages as phase-space integrals
\es{Wuse}{
\int d^nx \langle x_1,
\dots,x_n| \hat A| x_1,\dots, x_n\rangle=\int\frac{\prod_{i=1}^n dp_idx_i}{(2\pi\hbar)^n}(\hat A)_W\,.
}
We can apply it to the grand canonical average in \eqref{traceGC} to get
\es{Wav}{
\langle\cO_n\rangle^\text{GC}(\mu)=\Xi\int\frac{\prod_{i=1}^n dp_idx_i}{(2\pi\hbar)^n}(P_n\cO_n)_W\prod_{i=1}^n(\hat\rho_\text{GC})_W(x_i,p_i)\,,
}
where we integrated by parts $n$ times to remove the $\star$ in the products of $\cO_n$ and $\hat\rho_\text{GC}$, and we have chosen to have the antisymmetrizer act on the states in the Wigner transform \eqref{antiW} of $\cO_n$, since the density matrix is usually a more complicated operator.

The Wigner transform of $\hat\rho_\text{GC}(\hat H)$ involves two distinct small $\hbar$ expansions. The first arises from taking the Wigner transform of a nontrivial function of $\hat H$.  This can be computed using the so-called Wigner-Kirkwood expansion, which expresses the Wigner transform of any function $f(\hat H)$ as
\es{WK}{
f(\hat H)_W=\sum_{r\geq0}\frac{f^{(r)}(H_W)}{r!}\mathcal{G}_r\,,\qquad \mathcal{G}_r=\left[\left(\hat H-H_W(x,p)\right)^r\right]_W\,.
}
The lowest couple Wigner-Kirkwood coefficients $\mathcal{G}_r$ are trivially
\es{Gtrivial}{
\mathcal{G}_0=1\,,\qquad \mathcal{G}_1=0\,,
}
while $\mathcal{G}_r$ for $r\geq2$ can be computed using the product rule \eqref{wigProd} and have an $\hbar$ expansion
\es{Gnontrivial}{
\mathcal{G}_r=\sum_{n\geq\left[\frac{r+2}{3}\right]}\hbar^{2n}\mathcal{G}_r^{(n)}\,,\qquad r\geq2\,.
}
For instance, the lowest couple are
\es{kirks}{
\mathcal{G}_2&=-\frac{\hbar^2}{4}\left[\frac{\partial^2H_W}{\partial x^2}\frac{\partial^2H_W}{\partial p^2}-\left(\frac{\partial^2H_W}{\partial x\partial p}\right)^2 \right]+O(\hbar^4)\,,\\
\mathcal{G}_3&=-\frac{\hbar^2}{4}\left[\left(\frac{\partial H_W}{\partial x}\right)^2\frac{\partial^2H_W}{\partial p^2}+\left(\frac{\partial H_W}{\partial p}\right)^2\frac{\partial^2H_W}{\partial x^2}-2\frac{\partial H_W}{\partial x}\frac{\partial H_W}{\partial p}\frac{\partial^2 H_W}{\partial x\partial p} \right]+O(\hbar^4)\,,
}
and we show higher orders in Appendix \ref{hbarCorrApp}. The second $\hbar$ expansion involved in the computation of $(\hat\rho_\text{GC}(\hat H))_W$ comes from computing the Wigner transform of $\hat H$ itself. Let us assume that the $\hat p$ and $\hat x$ dependence of $\hat H$ factorizes, as it does in all the matrix models we consider, so that the canonical density matrix $\hat\rho$ is
\es{Hsimple}{
\hat\rho=e^{-\hat H}\,,\qquad \hat H=U(\hat x)+T(\hat p)\,.
}
The Wigner transform of $\hat H$ is then computed as
\es{WigH}{
(\hat H)_W&=\log_\star (\hat\rho)_W\\
&=U(x)+T(p)+\frac{\hbar^2}{24}\left[U'(x)^2T''(p)-2T'(p)^2U''(x)\right]+O(\hbar^4)\,,
}
where $\log_\star$ gives an expansion in $\hbar^2$, and we show higher orders in Appendix \ref{hbarCorrApp}. 

To sum up, the calculation of the canonical average $\langle \cO_n\rangle$ of an $n$-body operator requires the following steps:
\begin{enumerate}
\item Compute the Wigner transform $(P_n \cO_n)_W$ of the anti-symmetrized operator using \eqref{P} and \eqref{antiW}.
\item Compute the Wigner transform $(\hat \rho_\text{GC}(\hat H))_W$ of the single particle grand canonical density matrix in a small $\hbar$ expansion using the Wigner-Kirkwood \eqref{WK} and quantum Hamiltonian \eqref{WigH} expansions.
\item Compute the $2n$ phase space integrals in \eqref{Wav} and multiply by the grand potential $\Xi$ to get the grand canonical average ${\langle \cO_n \rangle^\text{GC}(\mu)}$.
\item Recover the canonical average $\langle \cO_n\rangle$ by performing the contour integral over $\mu$ in \eqref{grand}.
\end{enumerate}
The grand potential $\Xi$ for the 3d $\mathcal{N}=4$ theories that are amenable to the Fermi gas technique generically take the form
\es{Xigen}{
\Xi(\mu)=e^{c_1\mu^3+c_2 \mu+c_3}\,,
}
where $c_i$ are various theory-dependent constants. One can perform the contour integral in \eqref{grand} and use the identity
\es{Airystat}{
\text{Ai}(z)=\frac{1}{2\pi i }\int d\mu e^{\frac{\mu^3}{3}-z\mu}\,,
}
to get the famous Airy function behavior of the partition function $Z$. The $n$-body operators that we consider will all have grand canonical averages that are polynomial in $\mu$:
\es{ourGC}{
\langle \cO_n \rangle^\text{GC}(\mu)=\sum_{i=0}^{2n}b_i(\hbar)\mu^i\,,
}
where the coefficients $b_i(\hbar)$ were computed in a small $\hbar$ expansion. From the generic form of $\Xi$ in \eqref{Xigen} and the identity
\es{airtyDer}{
\text{Ai}''(z)=z\text{Ai}(z)\,,
}
we see that the $N$-dependence of the expectation values of $\langle \cO_n\rangle$ will be completely captured by $\text{Ai}(x)$ and $\text{Ai}'(x)$. We will demonstrate this for specific models in the following sections.

\section{ABJM matrix model}
\label{abjm}

We will start by discussing $n$-body operators in the $\mathcal{N}=6$ $U(N)_k\times U(N)_{-k}$ ABJM theory. These theories describe $N$ M2 branes probing a $\mC^4 /\mZ_{k}$ singularity and for finite $k$ the holographic dual is described by M-theory on $AdS_4\times S^7/\mZ_{k}$. If we parametrize $S^7$ by complex coordinates on $\mC^4$ subject to $\sum_{i=1}^4|z_i|^2=1$, then the $\mZ_{k}$ acts as
\es{kM2}{
\{z_1,z_2,z_3,z_4\}\to e^{2\pi i \over k}\{ z_1,z_2,z_3 ,z_4\}\,.
}
In the large $N$ and large $k$ limit with fixed $N/k$ the bulk description is Type IIA string theory on $AdS_4\times \mathbb{CP}^3$ \cite{Aharony:2008ug}.

First we will review the matrix model of this theory, and discuss how certain OPE coefficients of protected operators can be computed either as the expectation values of 2 and 4 body operators in the matrix model, or by taking mass derivatives of the mass deformed matrix model partition function. The latter quantity was previously computed for finite mass in terms of Airy functions. We will rederive this result using our methods by computing the grand canonical averages of these 2 and 4 body operators in a small $\hbar=2\pi k$ expansion to several orders, which yields the canonical averages written as Airy functions that capture the full $1/N$ expansion to the lowest few orders in $k$. This matches the known result and serves as a check on our methods.

\subsection{Matrix model}
\label{modelabjm}

ABJM theory is an $\mathcal{N}=6$ SCFT, so it has an $SO(6)_R\cong SU(2)_R\times SU(2)_R\times U(1)_R$ R-symmetry as well as a $U(1)$ flavor symmetry. In $\mathcal{N}=2$ language, ABJM theory consists of vector multiplets for each $U(N)_k\times U(N)_{-k}$ gauge group, as well as four chiral multiplets $Z^a,W_a$ for $a=1,2$ which transform under the gauge groups and the $SU(2)_R\times SU(2)_R\times U(1)$ flavor symmetry as shown in Table \ref{tab1}. 

\begin{table}[htpbp]
\begin{center}
\begin{tabular}{|l|c|c|c|}
\hline
 field     & $U(N)\times U(N)$    & $SU(2)\times SU(2)$             & $U(1)$  \\
 \hline 
 $Z^a$     & $(\overline{\bf N},{\bf N})$    & $({\bf 2},{\bf 1})$ & $1$   \\
 $W_a$  & $({\bf N},\overline{\bf N})$     &  $({\bf 1},\overline{\bf 2})$   & $-1$    \\
  \hline
\end{tabular}
\end{center}
\caption{Matter content of ABJM theory and their transformation under gauge and flavor symmetries in $\mathcal{N}=2$ language.}
\label{tab1}
\end{table}

The partition function on a squashed sphere with squashing parameter $b$ and deformed by masses $m_i$ for the chiral fields can then be computed by assembling the standard $\mathcal{N}=2$ ingredients, as reviewed in \cite{Willett:2016adv}, to get 
 \es{PartFunc}{
  Z_b(m_1, m_2, m_3) &= \frac{1}{(N!)^2}
   \int d^N \lambda \, d^N \mu\, e^{i \pi k \left[\sum_i \lambda_i^2 - \sum_j \mu_j^2 \right]} \prod_{i>j} 4 \sinh \left[\pi b(\lambda_i - \lambda_j)  \right] \sinh \left[\pi b^{-1}(\lambda_i - \lambda_j)\right] \\
   &{}\times \prod_{i>j} 4 \sinh \left[\pi b(\mu_i - \mu_j)  \right] \sinh \left[\pi b^{-1}(\mu_i - \mu_j)\right] 
    \prod_{i, j} \biggl[s_b\left(\frac{i Q}{4} - (\lambda_i - \mu_j + \frac{m_1 + m_2 + m_3}{2}) \right) \\
    &{}\times s_b\left(\frac{i Q}{4} - (\lambda_i - \mu_j + \frac{m_1 - m_2 - m_3}{2} ) \right) 
     s_b\left(\frac{i Q}{4} - (-\lambda_i + \mu_j +\frac{-m_1 - m_2 + m_3}{2}) \right) \\
     &{}\times s_b\left(\frac{i Q}{4} - (-\lambda_i + \mu_j + \frac{-m_1 + m_2 - m_3}{2}) \right) \biggr]\,,
 }
where $Q=b+\frac 1b$, the $\lambda_i,\mu_i$ correspond to the Cartans of the two gauge fields, and each chiral field contributes a factor with masses $m_i$ determined by the charge assignments in Table \ref{tab1}, so that $m_1$ corresponds to $U(1)$ and $m_2+m_3,m_2-m_3$ correspond to the Cartans of each factor in $SU(2)\times SU(2)$, respectively. The functions $s_b(x)$ are reviewed in Appendix \ref{special}. If we restrict to the round sphere with $b=1$, and set $m_3$ (or $m_2$) zero, then we can simplify the partition function using identities in \ref{special} and write it in terms of $m_\pm=m_2\pm m_1$ (or $m_\pm=m_3\pm m_1$) as
 \es{ZABJM}{
 & Z({m_+,m_-} )= \frac{1}{N!^2} \int d^N \lambda\, d^N \mu\, e^{i \pi k \left[ \sum_i \lambda_i^2 - \sum_j \mu_j^2 \right]} \\
  &\times \frac{\prod_{i<j} \left( 4 \sinh^2 \left[ \pi (\lambda_i - \lambda_j) \right] \right) 
    \prod_{i<j} \left( 4 \sinh^2 \left[ \pi (\mu_i - \mu_j) \right] \right) }{\prod_{i, j} \left(
     4 \cosh \left[\pi (\lambda_i - \mu_j + m_+/2  ) \right] 
     \cosh \left[\pi (\mu_i - \lambda_j + m_-/2  ) \right]  \right) } \,.
 }
As shown in \cite{Kapustin:2010xq}, the partition function can furthermore be simplified using the Cauchy determinant formula to take the form 
 \es{ZABJM2}{
 Z(m_+,m_-)= \frac{1}{N!}\int \frac{d^Nx}{(8\pi k)^N}\frac{\prod_ie^{\frac{m_-}{2}i  x_i}\sech \frac{ x_i}{2}\prod_{i<j}\sinh^2\frac{ x_{ij}}{2k}}{\prod_{i,j}\cosh(\frac{ x_{ij}}{2k}+\frac{ \pi m_+}{2})}\,,
 }
 where we integrated out $\mu_i$ and rescaled $\lambda_i\to x_i/(2\pi k)$. For $k=1$, this is the matrix model for a $U(N)$ gauge theory with an adjoint hypermultiplet and a fundamental hypermultiplet, deformed by an adjoint mass $m_+$ and an FI parameter $\frac{m_-}{2}$, which is expected from mirror symmetry \cite{Kapustin:2010xq}.\footnote{A similar single eigenvalue partition function can be written for $k=1$ and finite $b$ using $s_b(x)$ functions, as was shown numerically for the $m_{\pm}=0$ case in \cite{Hatsuda:2016uqa}.} For $k>1$, this mathematical reformulation has no physical explanation. Lastly, as shown in \cite{Nosaka:2015iiw}, one can use the Cauchy determinant formula to further write this partition function (up to an unimportant overall numerical factor) in the form of \eqref{Zintro} with single particle density matrix 
\es{rhoABJM}{
\hat\rho(m_+,m_-)=\frac{e^{\frac{im_+\hat p}{2}}}{2\cosh\frac{\hat p}{2}}\frac{e^{\frac{im_-\hat x}{2}}}{2\cosh\frac{\hat x}{2}}\,,
}
where the canonical variables are normalized so that
\es{hbarABJM}{
\hbar=2\pi k\,.
}
The grand partition function $\Xi$ was then computed in \cite{Nosaka:2015iiw} using the standard Fermi gas methods to all orders in $\hbar$ to get
\es{xiABJM}{
\Xi(m_+,m_-)&=e^{\frac C3\mu^3+B\mu+A}+O(e^{-\mu})\,,\\
C &= \frac{2}{\pi^2 k (1 + m_+^2) (1 + m_-^2)} \,, \qquad
   B = \frac{\pi^2 C}{3} - \frac{1}{6k} \left[ \frac{1}{1 + m_+^2} + \frac{1}{1 + m_-^2} \right] + \frac{k}{24} \,, \\
  A&= \frac{{\cal A}[k(1 + i m_+)] + {\cal A}[k(1 - i m_+)] +  {\cal A}[k(1 + i m_-)] + {\cal A}[k(1 - i m_-)] }{4}  \,,
}
 where the constant map function ${\cal A}$ is given by \cite{Hanada:2012si}
\es{constantMap}{
{\cal A}(k)&=\frac{2\zeta(3)}{\pi^2k}\left(1-\frac{k^3}{16}\right)+\frac{k^2}{\pi^2}\int_0^\infty dx\frac{x}{e^{kx}-1}\log\left(1-e^{-2x}\right)\\
&=\frac{2 \zeta (3)}{\pi ^2 k}-\frac{k}{12}+\sum_{n=1}^\infty k^{n+2}\frac{B_{n+1}  \text{Li}_{n+3}(1)}{\pi ^22^{n+2}}\,,
}
 and in the second line we wrote $\cA$ in the small $k$ expansion, where $B_{n+1}$ is a Bernoulli number. The mass deformed partition function $Z(m_+,m_-)$ can then be obtained by computing the $\mu$ contour integral in \eqref{xi} to get
 \es{GotZABJM}{
  & Z = e^A C^{-\frac 13} \text{Ai}\left[C^{-\frac 13} (N-B) \right]+(\text{non-perturbative in $N$}) \,,\\
 } 
 where the non-perturbative corrections come from both the $e^{-\mu}$ terms that were left undetermined in \eqref{xiABJM}, which translate into $e^{-\sqrt{N k}}$ terms, as well as $e^{-\sqrt{N/k}}$ terms that are invisible in the small $k$ expansion of the Fermi gas method.
 
 As discussed in the Introduction, taking $n$ derivatives of the free energy $\partial_{m_\pm}^n\log Z(m_+,m_-)\big\vert_{m=0}$ for ABJM gives $n$-point functions of stress tensor multiplet operators integrated on the sphere $S^3$. For instance, taking two derivatives we find that $c_T$ in the normalization of \cite{Agmon:2017xes} is
 \es{cTABJM}{
c_T=&-\frac{64}{\pi^2}{\partial^2_{m_\pm}\log Z}\big\vert_{m_\pm=0}\\
=&\frac{32k^2{\cal A}''[k]}{\pi^2}-\frac{128}{3\pi^2}+\frac{8\cdot 2^{\frac23}(k^2-24kN-28)\text{Ai}'\left[-\frac{(8+k^2-24kN)\pi^{\frac23}}{24\cdot 2^{\frac13}k^{\frac23}}\right]}{9k^{\frac23}\pi^{\frac43}\text{Ai}\left[-\frac{(8+k^2-24kN)\pi^{\frac23}}{24\cdot 2^{\frac13}k^{\frac23}}\right]}+\text{non-perturbative}\,.
 }
Taking four derivatives, we get the integrated stress tensor multiplet four point function. As discussed in \cite{Agmon:2017xes,Binder:2019mpb}, this quantity can be considered in the 1d protected subsector, where only a finite number of protected operators appear. It can then be used to compute the OPE coefficient of a certain protected operator in the $\bold{84}$ of $SO(6)_R$ as
\es{84}{
\lambda_{\bold{84}}^2=&2+\frac{2^{12}}{\pi^4 c_T^2}{\partial^4_{m_\pm}\log Z}\big\vert_{m_\pm=0}\,,
}
which from \eqref{GotZABJM} can also be written to all orders in $1/N$ as a complicated sum of Airy functions and their derivative, similar to $c_T$, whose explicit form we will not use. 

Instead of computing these mass derivatives using the all orders in $1/N$ Airy function expression for the mass deformed partition function in \eqref{GotZABJM}, one could write them as $n$-body operators in the $m_\pm=0$ matrix model with Hamiltonian \eqref{Hsimple}
\es{mass0H}{
\hat H=U(\hat x)+T(\hat p)\,,\qquad U(x)=\log(2\cosh\frac x2)\,,\qquad T(p)=\log(2\cosh\frac p2)\,,
}
by taking derivatives of the exact mass deformed partition function \eqref{ZABJM2}. In particular, we find that ${\partial^2_{m_\pm} Z}\big\vert_{m_\pm=0}$ can be expressed as $(n\leq2)$-body operators
 \es{pmOps}{
   {\partial^2_{m_+} Z}\big\vert_{m_\pm=0}&=\langle\cO^+\rangle\equiv\langle\cO^+_0+\cO^+_2\rangle\,,\qquad\;\, \cO^+_0=-\frac{\pi^2}{4}N\,,\qquad\quad\;\cO^+_2=-\frac{\pi^2}{4}\sum_{i\neq j}\sech^2\frac{ x_{ij}}{2k}\,,\\
{\partial^2_{m_-} Z}\big\vert_{m_\pm=0}&=\langle\cO^-\rangle\equiv\langle\cO^-_1+\cO^-_2\rangle\,,\qquad\;\, \cO^-_1=-\frac{1}{4}\sum_i x_i^2\,,\qquad \cO^-_2=-\frac14\sum_{i\neq j}x_ix_j \,.
 }
 Similarly, we can express ${\partial^4_{m_-} Z}\big\vert_{m_\pm=0}$ as $(n\leq4)$-body operators:
 \es{O4}{
{\partial^4_{m_-} Z}\big\vert_{m_\pm=0}=\langle\cO^4\rangle\equiv\sum_{a=1}^4\langle\cO_a^4\rangle\,,\qquad \cO^4_1=&\frac{1}{16} \sum_{i}x_i^4\,, \qquad\qquad \;\; \cO^4_2=\frac{1}{16} \sum_{i\neq j}(x_i^3x_j+x_i^2x_j^2) \,, \\
\cO^4_3=&\frac{1}{16} \sum_{i\neq j\neq k}x_i^2x_jx_k \,,\qquad \cO_4^4=\frac{1}{16} \sum_{i\neq j\neq k\neq l}x_ix_jx_kx_l\,,
 }
 as well as a similar more complicated operator from taking four derivatives of $m_+$. From \eqref{xiABJM}, we expect that the grand canonical averages of these operators in the $m_\pm=0$ matrix model should be
 \es{expectation2}{
 \langle \cO^\pm\rangle^\text{GC}=&-\frac{4}{3\pi^2k}\mu^3-\frac{1}{k}\mu-\frac{k^2\mc{A}^{\prime\prime}(k)}{2}+O(e^{-\mu})\,,\\
 }
 for the 2-body operators and
 \es{expectation4}{
   \langle \cO^4\rangle^\text{GC}=&\frac{16}{3\pi^4k^2}\mu^6+\frac{8}{\pi^2k^2}\mu^4+\frac{4\left(k^2\mathcal{A}''(k)+4\right)}{\pi^2k}\mu^3+\frac{3}{k^2}\mu^2\\
   &+\frac{3\left(k^2\mathcal{A}''(k)+4\right)}{k}\mu+\frac{1}{4}k^4\left(2\mathcal{A}^{(4)}(k)+3\mathcal{A}''(k)^2\right)\,,\\
 } 
for the 4-body operator, where the small $k$ expansion of $\mathcal{A}(k)$ is given in \eqref{constantMap}. In the next few subsections we will use our $n$-body formalism to recover all the polynomial in $\mu$ terms shown here as well as the constant term to the lowest few orders in $k$. This will test all aspects of our formalism, which we will then use in the next section to compute completely new physical data.

\subsection{$\partial_{m_-}^2 Z\big\vert_{m=0}$}
\label{mm}

We start by taking the Wigner transform \eqref{antiW} of the antisymmetrized one- and two-body operators for $\cO_-$ in \eqref{pmOps} in their respective $n$-particle Hilbert spaces to get
\es{wigOm}{
(\cO^-_1)_W=-\frac14x^2\,,\quad (P_2\cO^-_2)_W=-\frac14\left[ x_1x_2-2\pi \hbar x_1^2\delta(x_{12})\delta(p_{12})-\pi\hbar^3\delta(x_{12})\delta''(p_{12})\right]\,,
}
where $x_{ij}\equiv x_i-x_j$ and $p_{ij}\equiv p_i-p_j$. We then compute the phase space integral \eqref{Wav} for both operators to get
\es{intOm}{
\frac{\langle \cO^-\rangle^\text{GC}}{\Xi}=\text{A}+\text{B}\,,\qquad \text{A}\equiv\int \frac{dxdp}{2\pi\hbar}\frac{x^2}{4}\rho_W(\rho_W-1)\,,\qquad \text{B}\equiv\int\frac{dxdp}{2\pi\hbar}\frac{\hbar^2}{8}\left(\rho_W\partial_p^2\rho_W-(\partial_p\rho_W)^2\right)\,,
}
where we denote $\rho_{W}\equiv(\hat\rho_\text{GC}(\mu))_W(x,p)$, the leading term in $(P_2\cO^-_2)_W$ vanished since it is odd in each variable, and the delta functions allowed us to simplify the 2-body integrals. We will be able to compute the $\hbar$ expansion of the first term A analytically, while we will need to use numerics for the second term B. The reason A can be computed analytically is that the products of $\rho_\text{GC}(\mu)=\frac{1}{e^{H-\mu}+1}$ can be simplified as
\es{rhosimp}{
\rho_\text{GC}(\rho_\text{GC}-1)=-\partial_\mu \rho_\text{GC}=-\pi\partial^2_\mu \csc(\pi \partial_\mu)\theta(\mu- H)\,,
}
where in the second equality we used the Sommerfeld expansion as shown in \cite{Klemm:2012ii} to write $\rho$ in terms of derivatives of the Heaviside function $\theta$. We can then apply the Wigner-Kirkwood expansion to the simpler function $\theta(\mu-\hat H)$ to express A as
\es{rhosimp2}{
\text{A} =-\pi\partial^2_\mu \csc(\pi \partial_\mu)\int \frac{dxdp}{2\pi\hbar}\frac{x^2}{4}\left[\theta(\mu-H_W)+\sum_{r=2}^\infty\frac{\mathcal{G}_r}{r!}\delta^{(r-1)}(\mu-H_W)\right]\,,
}
where recall that the coefficients $\mathcal{G}_r$ have an $\hbar$ expansion given in \eqref{Gnontrivial}, and the quantum Hamiltonian $H_W$ also has an $\hbar$ expansion \eqref{HamApp}. The resulting phase space integrals are very similar to those performed in the original Fermi gas paper \cite{Marino:2011eh}, and as shown in Appendix \ref{phase} yield
\es{Om1}{
\text{A} =-\frac{1}{k}\left(\frac{4\mu^3}{3\pi^2}+\mu+\frac{2\zeta(3)}{\pi^2}\right)+
\frac{1}{12}k(\mu-1)-
\frac{\pi^2k^3}{1440}+
O(k^5)\,,
}
where we set $\hbar=2\pi k$. The leading $k$ terms indeed match the expectation in \eqref{expectation2}, where the constant map $\mathcal{A}(k)$ was expanded using \eqref{constantMap}.

For B, we cannot apply the Sommerfeld expansion due to the $p$ derivatives of the $\rho_W$ term. Instead, we apply the Wigner-Kirkwood expansion directly to each $\rho_W$ in \eqref{intOm} to get
\es{Bint}{
\text{B}\equiv\int\frac{dxdp}{2\pi\hbar}\frac{\hbar^2}{8}\sum_{r,s=0}^\infty\frac{\mathcal{G}_r\mathcal{G}_s}{r!s!}\left(\rho_\text{GC}^{(r)}(H_W)\partial_p^2\rho_\text{GC}^{(s)}(H_W)-\partial_p\rho_\text{GC}^{(r)}(H_W)\partial_p\rho_\text{GC}^{(s)}(H_W)\right)\,.
}
We can then expand the Wigner-Kirkwood coefficients and the quantum Hamiltonian in $\hbar$ to get two-dimensional integrals at each order in $\hbar$ that we write explicitly in Appendix \ref{phase}. These integrals involve derivatives of $\rho_\text{GC}$ that we were unable to evaluate analytically. Instead, we computed them numerically for many values of $\mu$ to high precision to get
\es{Bresult}{
\text{B}=-\frac{k}{12}(\mu-1)+k^3\frac{\pi^2}{720}+O(k^5)\,.
}
Finally, we combine A and B to find 
\es{OmFin}{
\frac{\langle\cO^-\rangle^\text{GC}}{\Xi} =-\frac{4}{3k\pi^2}\mu^3-\frac{1}{k}\mu-\frac{2\zeta(3)}{\pi^2k}+\frac{ \pi^2 k^3}{1440} +O(k^5)\,,
}
which matches the expectation \eqref{expectation2} to the order shown.

\subsection{$\partial_{m_+}^2 Z\big\vert_{m=0}$}
\label{mp}

Next, we consider the other two body operator $\cO^+$, which is expected to have the same expectation value as $\cO^-$. The integrals that appear in this calculation are more challenging than those of $\cO^-$, so we will work just to leading order in $\hbar$. The operator $\cO^+$ consists of a zero body operator $\cO_0^+$ whose expectation value is trivial, as well as the 2-body operator $\cO^+_2$ whose anti-symmetrized Wigner transform is 
\es{wigOp}{
(P_2\cO^+_2)_W=-\frac{\pi^2}{4}\left[\sech^2\frac{\pi x_{12}}{\hbar}-\frac{\hbar}{\pi}p_{12}\csch(\frac{p_{12}}{2})\delta(x_{12})\right]\,.
}
The phase space integral of $\cO^+_2$ is then
\es{intOp}{
\frac{\langle \cO^+_2\rangle^\text{GC}}{\Xi}&=-\frac{\pi^2}{4}\int \frac{d^2xd^2p}{(2\pi\hbar)^2}\rho_W(x_1,p_1)\rho_W(x_2,p_2)\left[\sech^2\frac{\pi x_{12}}{\hbar}-\frac{\hbar}{\pi}p_{12}\csch(\frac{p_{12}}{2})\delta(x_{12})\right]\\
&=-\frac{1}{8}\int \frac{dxd^2p}{2\pi\hbar} \rho_\text{GC}(x,p_1)\rho_\text{GC}(x,p_2) \left[2-p_{12}\csch(\frac{p_{12}}{2})\right]+O(\hbar)\,,
}
where in the second line we expanded $\sech^2(x)$ for large $x$ and replaced $\rho_W$ by its classical value to leading order in $\hbar$ . We then evaluate this integral numerically and set $\hbar=2\pi k$ to get
\es{OpFin}{
\frac{\langle\cO^+_2\rangle^\text{GC}}{\Xi} =-\frac{4}{3k\pi^2}\mu^3+\frac{\mu ^2}{2 k}-\frac{1}{k}\mu-\frac{2\zeta(3)}{\pi^2k}+\frac{\pi ^2}{12 k}+O(k)\,.
}
To compare to the expected grand canonical average \eqref{expectation2}, we need to add the zero-body $-\frac{\pi^2}{4}N$ term, but $N$ is strictly speaking not defined in the grand canonical ensemble. The physical quantity is the canonical ensemble average, which we get by taking the $\mu$ integral in  \eqref{grand} of \eqref{OpFin} and adding $-\frac{\pi^2}{4}N$ to get
\es{OpFinCan}{
\langle\cO^+\rangle &= -\frac{\pi^2}{4}N+\frac{1}{2\pi i}\int d\mu e^{-\mu N}\Xi(\mu)\langle \cO^+_2\rangle^\text{GC}(\mu)\\
& =\frac{1}{2\pi i}\int d\mu e^{-\mu N}\Xi(\mu)\left[\langle \cO^+_2\rangle^\text{GC}(\mu)-\frac{\pi^2}{4}\left(\frac{2 \mu ^2}{\pi ^2 k}+\frac{k}{24}+\frac{1}{3 k}\right)\right]\,,\\
}
where the second line followed from applying integration by parts using the definition of $\Xi$ in \eqref{xiABJM} for $m_\pm=0$. The term in the square brackets matches the expected grand canonical average in \eqref{expectation2} to leading order in $k$.

\subsection{$\partial_{m_-}^4 Z\big\vert_{m=0}$}
\label{mmmm}

Finally, we consider the expectation value of the $(n\leq4)$-body operator $\cO^4$ defined in \eqref{O4}, whose grand canonical expectation value is expected to take the form \eqref{expectation4}. As with the previous subsection, we will consider this calculation to leading order in $\hbar$. From taking the anti-symmetrized Wigner transforms of the operators in \eqref{O4}, we find that the leading order in $\hbar$ contribution comes from the terms
\es{wig4}{
(P_4\cO^4_4)_W=&\frac{3}{16}{(2\pi\hbar)^2} x_1^2x_3^2\delta(x_{12})\delta(x_{34})\delta(p_{12})\delta\left(p_{34}\right) +O(\hbar^4)\,,\\
(P_3\cO^4_3)_W=&- \frac{3}{8}2\pi\hbar x_1^2 x_2^2 \delta(x_{23})\delta(p_{23}) +O(\hbar^3)\,,\\
(P_2\cO^4_2)_W=&\frac{3}{16}x_1^2x_2^2+O(\hbar^2)\,.\\
} 
We use these to compute the leading order phase space integral
\es{intO4}{
\frac{\langle \cO^4\rangle^\text{GC}}{\Xi}&=\frac{3}{16}\left(\int \frac{dxdp}{2\pi\hbar}x^2 \rho_\text{GC}(x,p)(1-\rho_\text{GC}(x,p))\right)^2+O(\hbar^{-1})\,.\\
}
where we used the classical expressions for the density matrices. This quantity is proportional to the square of the classical expression for the term A in the phase space integrals for $\langle\cO^-\rangle^\text{GC}$ in \eqref{intOm}. We can thus simply square the result for A in \eqref{Om1} to get the leading order in $\hbar=2\pi k$ result
\es{O4Fin}{
\frac{\langle\cO^4\rangle^\text{GC}}{\Xi} =\frac{1}{k^2}\left(\frac{16}{3\pi^4}\mu^6+\frac{8}{\pi^2}\mu^4+\frac{16\zeta(3)}{\pi^4}\mu^3+3\mu^2+\frac{12\zeta(3)}{\pi^2}\mu+\frac{12\zeta(3)^2}{\pi^4}\right)+O\left(\frac{1}{k}\right)\,,
}
which matches \eqref{expectation4} to leading order. This concludes our calculations in the ABJM model, which served as checks of our new method against old results. 

\section{$N_f$ matrix model}
\label{nf}

We will now discuss $n$-body operators in the so-called $N_f$ matrix model introduced in \cite{Grassi:2014vwa} (see also \cite{Moriyama:2014gxa}), which is an $\mathcal{N}=4$ theory of a $U(N)$ vector multiplets coupled to one hypermultiplet in the adjoint representation and $N_f$ hypermultiplets in the fundamental. These theories describe $N$ M2 branes probing a $\mC^2\times \mC^2/\mZ_{N_f}$ singularity and for finite $N_f$ the holographic dual is described by M-theory on $AdS_4\times S^7/\mZ_{N_f}$. If we parametrize $S^7$ by complex coordinates on $\mC^4$ subject to $\sum_{i=1}^4|z_i|^2=1$, then the $\mZ_{N_f}$ acts as
\es{NfM2}{
\{z_1,z_2,z_3,z_4\}\to \{ z_1,z_2,z_3 e^{2\pi i \over N_f},z_4e^{-{2\pi i \over N_f}}\}
}
In other words, we have M2s probing a KK monopole of charge $N_f$. In the large $N$ and large $N_f$ limit with fixed $N/N_f$ the bulk description is Type IIA string theory, where the KK monopole is realized by $N_f$ D6 branes wrapping the fixed loci of $\mZ_{N_f}$, which is $AdS_4\times S^3$ \cite{Benini:2009qs,Bashkirov:2010kz}.

We will first review the matrix model of this theory, and discuss how $c_T$ can be computed from the matrix model partition function deformed by a squashing parameter $b$, by taking two derivatives of $b$ and setting $b=1$. This defines a 2-body operator that we will compute in a small $\hbar=2\pi $ expansion to several orders, which will end up corresponding to a small $N_f$ expansion. The corresponding canonical average can then written as Airy functions that capture the full $1/N$ expansion to the lowest few orders in $N_f$. Unlike the previously discussed ABJM cases, this is a new result. 

\subsection{Matrix model}
\label{modelnf}

The $N_f$ theory is an $\mathcal{N}=4$ $U(N)$ gauge theory with an $SU(2)\times SU(N_f)$ flavor symmetry. It has a vector multiplet for the $U(N)$ gauge symmetry, $N_f$ hypermultiplets in the fundamental of $U(N)$ and $SU(N_f)$, and a hypermultiplet in the adjoint of $U(N)$ and the fundamental of $SU(2)$. The partition function on a squashed sphere with squashing parameter $b$ and deformed by masses for the matter fields as well as a FI parameter for the gauge field can then be computed by assembling the standard $\mathcal{N}=4$ ingredients in e.g. \cite{Willett:2016adv} to get 
 \es{Znf}{
 Z_b(m_\text{adj},&m_\alpha,\zeta)= \frac{1}{N!}\int \frac{d^Nx}{(2^{N_f+1})^N}{\prod_ie^{\pi\zeta   ix_i}\prod_{i<j}\sinh({ \pi bx_{ij}})\sinh({\pi  x_{ij}}{b^{-1}})} \\
 &\times {\prod_{i,j}2s_b\Big(\frac{iQ}{4}+{ x_{ij}}+\frac{  m_\text{adj}}{2}\Big)s_b\Big(\frac{iQ}{4}-({ x_{ij}}+\frac{  m_\text{adj}}{2})\Big)}\\
 &\times\prod_i { \Big[2s_b\Big(\frac{iQ}{4}+{ x_i}+\sum_{\alpha=1}^{N_f-1}\frac{ m_{\alpha,i}}{2}\Big)s_b\Big(\frac{iQ}{4}-({ x_i}+\sum_{\alpha=1}^{N_f-1}\frac{ m_{\alpha,i}}{2})\Big)\Big]^{N_f}}\,,
 }
where $x_i$ correspond to the Cartans of the gauge fields, the $m_{\alpha,i}$ correspond to the $N_f-1$ Cartans of the flavor group $SU(N)$,  $m_\text{adj}$ corresponds to the Cartan of the $SU(2)$ flavor symmetry, and $\zeta$ is the FI parameter for the gauge group. The functions $s_b(x)$ are reviewed in Appendix \ref{special}. If we restrict to the round sphere with $b=1$, set $N_f=1$, and rescale $x_i\to x_i/(2\pi)$, then we can simplify the matrix model using the identities in Appendix \ref{special} so that it matches the ABJM matrix model in \eqref{ZABJM2} with $k=1$ and the identification $\zeta={m_-}$ and $m_\text{adj}=m_+$, as expected from mirror symmetry. For general $N_f$ but with $b=1$ and vanishing masses we find
 \es{Znf2}{
 Z= \frac{1}{N!}\int \frac{d^Nx}{( 2^{N_f+2}\pi)^N}{\prod_i\left(\sech \frac{ x_i}{2}\right)^{N_f}\prod_{i<j}\tanh^2\frac{ x_{ij}}{2}}\,.
 }
As shown in \cite{Grassi:2014vwa}, one can use the Cauchy determinant formula to further write this partition function (up to an unimportant overall numerical factor) in the form of \eqref{Zintro} with single particle density matrix 
\es{rhoNf}{
\hat\rho=\frac{1}{2\cosh\frac{\hat p}{2}}\frac{1}{(2\cosh\frac{\hat x}{2})^{N_f}}\,,
}
where the canonical variables are normalized so that
\es{hbarNf}{
\hbar=2\pi \,.
}
The Hamiltonian $\hat\rho=e^{-\hat H}$ can then be written as
\es{HamNf}{
\hat H=N_f U(\hat x)+T(\hat p)\,,\qquad U(x)=\log(2\cosh\frac x2)\,,\qquad T(p)=\log(2\cosh\frac p2)\,,
}
where these are the same $U(x)$ and $T(p)$ that appeared in the ABJM Hamiltonian \eqref{mass0H}. Unlike ABJM, for the $N_f$ model $\hbar$ does not depend on any physical parameters, so the small $\hbar$ expansion would seem to be unphysical. In practice, however, one finds that the polynomial in $\mu$ terms in grand canonical averages only receive a finite number of $\hbar$ corrections, just as in the ABJM case, so the small $\hbar$ expansion still serves as a useful way of deriving the Airy function behavior of matrix model observables, which captures the all orders in $1/N$ expansion. Furthermore, the $\hbar$ corrections tend to correlate with $N_f$ corrections. For instance, the grand partition function $\Xi$ was computed in \cite{Grassi:2014vwa,Hatsuda:2014vsa} using the standard Fermi gas methods to all orders in $\hbar$ to get
\es{xiNf}{
\Xi&=e^{\frac C3\mu^3+B\mu+A}+O(e^{-\mu})\,,\\
C &= \frac{2}{\pi^2 N_f} \,, \qquad
   B = \frac{1}{2N_f}- \frac{N_f}{8} \,, \qquad A= \frac{{\cal A}[N_f] + {\cal A}[1]N_f^2  }{2}  \,,
}
 where this is the same constant map function ${\cal A}$ that was defined for ABJM in \eqref{constantMap}. For $N_f=1$, this matches the ABJM case \eqref{xiABJM} for $m_\pm=0$ and $k=1$ as expected. The partition function $Z$ can then be obtained by computing the $\mu$ contour integral in \eqref{xi} to get the same expression \eqref{GotZABJM} as in the ABJM case, except with $A,B,C$ now given in \eqref{xiNf}.
 
 We can now consider taking derivatives of the various deformation parameters in \eqref{Znf}, as we did for the masses in ABJM. Derivatives of $m_{\alpha,i}$ or $m_\text{adj}$ will give integrated correlators of the currents for the $SU(N)$ or $SU(2)$ flavor symmetries, respectively, which are unrelated to the stress tensor multiplet since the $N_f$ theory only has $\mathcal{N}=4$ supersymmetry. In fact, the Fermi gas formalism can be applied to this model for finite masses, just as with the ABJM case, so there is no need to use the new $n$-body formalism in this work for mass derivatives. To compute $c_T$, however, one must take two derivatives of the squashing parameter $b$ as in \eqref{cTGen}, and the Fermi gas formalism cannot be applied to \eqref{Znf} for finite $b$.\footnote{An exception is the specific values $b^2=3$ or $b^2=1/3$, as shown in \cite{Hatsuda:2016uqa}.} Instead, we will apply our formalism to the $(n\leq2)$-body operator we derive from \eqref{Znf}:
 \es{Nfop}{
 \partial_b^2 Z\big\vert_{b=1} & =\langle\cO^b\rangle\equiv\sum_{a=0}^3\langle\cO^b_a\rangle\,,\qquad \cO^b_0=-\frac{\pi^2}{4}N\,,\\
  \cO^b_1&= -\frac{N_f}{4}\sum_{i}\left(x_i^2+x_i\sinh x_i+\pi^2\right)\sech^2\left(\frac{x_i}{2}\right)\,,\\
\cO^b_2&=\frac12\sum_{i\neq j}\csch{x_{ij}}\left(-\left(2x_{ij}^2+\pi^2\right)\coth x_{ij}+2x_{ij}+\pi^2\csch{x_{ij}}\right)\,.
}
 In the next subsection we will use our $n$-body formalism to compute all the $\hbar$ corrections to the polynomial in $\mu$ terms in the grand canonical averages of these expressions, as well as the constant term to the lowest few orders in $\hbar$, which will correspond to the lowest few orders in $N_f$.

\subsection{$\partial_{b}^2 Z\big\vert_{b=1}$}
\label{nfcalc}

The operator $\cO^b$ consists of a zero body operator $\cO_0^b$ whose expectation value is trivial, as well as the 1-body $\cO^b_1$ and 2-body $\cO^b_2$ operators whose anti-symmetrized Wigner transforms in their respective $n$-body Hilbert spaces are 
\es{wigOpb}{
(\cO^b_1)_W&= -\frac{N_f}{4}\left(x^2+x\sinh x+\pi^2\right)\sech^2\left(\frac{x}{2}\right)\,,\\
(P_2\cO^b_2)_W&= \frac12 \left({\csch}\left(x_{12}\right)\left(2x_{12}-\left(2x_{12}^2+\pi^2\right)\coth\left(x_{12}\right)+\pi^2{\csch}\left(x_{12}\right)\right)\right)\\
& \qquad+\delta\left(x_{12}\right)\frac{\pi ^2 \left((p_{12}) {\csch}\left(\frac{p_{12}}{2}\right)+1\right)}{\cosh \left(\frac{p_{12}}{2}\right)+1} \,.
}
The phase space integrals for $\cO^b_1$ and $\cO^b_2$ are then
\es{intOb}{
&\frac{\langle \cO^b_1\rangle^\text{GC}}{\Xi}=-\frac{N_f}{4}\int \frac{dxdp}{2\pi\hbar}\sum_{r=0}\frac{\mathcal{G}_r}{r!}\rho^{(r)}_\text{GC}(H_W(x,p))\left(x^2+x\sinh x+\pi^2\right)\sech^2\left(\frac{x}{2}\right)\,,\\
&\frac{\langle \cO^b_2\rangle^\text{GC}}{\Xi}=\int \frac{d^2xd^2p}{(2\pi\hbar)^2}\sum_{r,s=0}\frac{\mathcal{G}_r\mathcal{G}_s}{r!s!}\rho^{(r)}_\text{GC}(H_W(x_1,p_1))\rho^{(s)}_\text{GC}(H_W(x_2,p_2))\\
&\times\left[  \frac12 \left({\csch}\left(x_{12}\right)\left(2x_{12}-\left(2x_{12}^2+\pi^2\right)\coth\left(x_{12}\right)+\pi^2{\csch}\left(x_{12}\right)\right)\right)+\delta\left(x_{12}\right)\frac{\pi ^2 \left(p_{12} {\csch}\left(\frac{p_{12}}{2}\right)+1\right)}{\cosh \left(\frac{p_{12}}{2}\right)+1}  \right]\,,\\
}
where we applied the Wigner-Kirkwood expansion to the densities. We can then expand the Wigner-Kirkwood coefficients and the quantum Hamiltonian in $\hbar$ for the $n$-body operators to get $2n$-dimensional integrals at each order in $\hbar$ that we write explicitly in Appendix \ref{phase}. We computed these numerically for many values of $\mu$ to high precision to get\footnote{For the polynomial in $\mu$ terms, some parts of the integrals can be done analytically, and so their numerical precision is much higher. In contrast, some of the $\mu$-independent terms must be done completely numerically, and so it is harder to calculate them to the same precision.}
\es{ObFin}{
\frac{\langle\cO^b_1+\cO^b_2\rangle^\text{GC}}{\Xi} =-\frac{8}{3\pi^2 N_f}\mu^3+\frac{1}{2N_f}\mu^2-\left(\frac{5}{3N_f}+\frac{N_f}{3}\right)+.46 N_f^2-.55N_f+\frac{.56}{N_f}+O(N_f^{-2})\,,
}
where we set $\hbar=2\pi$ and we were unable to guess analytic formulae for the constant in $\mu$ terms. Note that just as in the ABJM case, the polynomial in $\mu$ terms only received a finite number of $\hbar$ corrections, while the constant in $\mu$ term receives $\hbar$ corrections to all orders. We can now get the canonical average $c_T=\langle \cO^b\rangle$ by taking the $\mu$ integral in \eqref{grand} and adding the zero-body term $-\frac{\pi^2}{4}N$ to get
\es{ObFinCan}{
c_T=\langle\cO^b\rangle &= -\frac{\pi^2}{4}N+\frac{1}{2\pi i}\int d\mu e^{-\mu N}\Xi(\mu)\langle \cO^b_1+\cO^b_2\rangle^\text{GC}(\mu)\\
& = -\frac{8 (2 \pi )^{2/3} \left(8 N N_f+3
   N_f^2+6\right) \text{Ai}'\left(\frac{\left(N_f^2+8 N
   N_f-4\right) \pi ^{2/3}}{8 \sqrt[3]{2}
   N_f^{2/3}}\right)}{3\pi^2N_f^{2/3} \text{Ai}\left(\frac{\left(N_f^2+8
   n N_f-4\right) \pi ^{2/3}}{8 \sqrt[3]{2} N_f^{2/3}}\right)}+\mathfrak{c}-\frac{128}{3\pi^2}+\text{non-perturbative}\,,
   }
where we performed the $\mu$ integral as in \eqref{Airystat} using the definition of $\Xi$ in \eqref{xiNf}, and the constant term $\mathfrak{c}$ was determined to the first couple orders in $N_f$:
\es{ANf}{
\mathfrak{c}=.46N_f^2-.24N_f-\frac{.67}{N_f}+O(N_f^{-2})\,.
}
If we set $N_f=1$ in \eqref{ObFinCan}, then we get the ABJM value of $c_T$ in \eqref{cTABJM} as expected, up to the constant term $\mathfrak{c}$ that is not yet completely fixed. We can also expand \eqref{ObFinCan} at large $N$ to get
\es{cTNflargeN}{
c_T=\frac{64 \sqrt{2} N^{3/2} \sqrt{{N_f}}}{3 \pi }+\frac{4 \sqrt{2} \sqrt{N} \left(7
   {N_f}^2+8\right)}{3 \pi  \sqrt{{N_f}}}+O(N^0)\,,
}
where note that these first two terms do not depend on $\mathfrak{c}$. The leading term in fact matches the leading large $N$ term for the ABJM $c_T$ if we identify $N_f=k$, which is expected since to leading order in Newton's constant  supergravity on $AdS_4\times S^7/\mathbb{Z}_k$ with $\mathbb{Z}_k$ action \eqref{kM2} is the same as $AdS_4\times S^7/\mathbb{Z}_{N_f}$ with $\mathbb{Z}_{N_f}$ action \eqref{NfM2},

\section{Conclusion}
\label{conc}

In this work we showed how OPE coefficients of protected operators in 3d $\mathcal{N}=4$ gauge theories with rank $N$ can be computed as statistical averages of $n$-body operators in a free Fermi gas, which can be computed to all orders in $1/N$ in terms of Airy functions using the WKB expansion. This generalizes the Fermi gas formalism originally derived in \cite{Marino:2011eh} to local CFT data, unlike previous studies which had only considered non-local quantities such as the sphere partition function and Wilson loops expectation values. We demonstrated our formalism in the case of ABJM theory, where we used it to compute various OPE coefficients that could be written as derivatives of the mass deformed sphere partition function $Z(m)$ at $m=0$. Since $Z(m)$ had already been computed for finite $m$ in \cite{Nosaka:2015iiw} using the Fermi gas formalism, our results reproduced known quantities. We then applied our formalism to the $U(N)$ gauge theory with one adjoint and $N_f$ fundamental hypermultiplets, in which case we derived a new all orders in $1/N$ result at finite $N_f$ up to an $N$-independent constant $\mathfrak{c}$ for the coefficient $c_T$ of the stress tensor two point function, which is inversely related to the OPE coefficient of two stress tensors and any scalar operator. The all orders in $1/N$ result is shown in \eqref{ObFinCan}, while the first two orders at large $N$, which are independent of $\mathfrak{c}$, are shown in \eqref{cTNflargeN}. It would be nice to compare this to a bulk calculation in the future, or to the 3d $\mathcal{N}=4$ numerical bootstrap results of \cite{Chang:2019dzt}. It would also be useful to generalize the lowest few orders in $N_f$ numerical result for $\mathfrak{c}$ shown in \eqref{ANf} to an all orders analytic result, perhaps in terms of the constant map function \eqref{constantMap} as was shown in the related context of the sphere free energy \cite{Hatsuda:2014vsa}.

The Fermi gas calculations in this work resemble those of the previous work \cite{Klemm:2012ii} on expectation values of Wilson loops in the fundamental representation, which could be written as averages of one-body operators in the Fermi gas, but are much simpler. While the grand canonical averages of Wilson loops were exponential in the chemical potential $\mu$, so that all orders in the WKB expansion were required to derive the Airy function behavior, the grand canonical averages of the $n$-body operators that appear in our formalism always take the form of polynomials in $\mu$ of finite degree, so only a finite number of terms need be computed to derive the all orders in $1/N$ Airy function behavior. Furthermore, in many cases the coefficients of the polynomial in $\mu$ terms receive only a few perturbative corrections in the WKB expansion, as was previously observed in Fermi gas calculations of the sphere free energy \cite{Marino:2011eh}. On the other hand, a technical difficulty that occurs in the case of $n$-body operators for $n>1$ is that there is no Sommerfeld-like expansion of products of Fermi distributions in a large $\mu$ expansion that can be used to simplify the phase space integrals, as was shown for one-body operators in \cite{Klemm:2012ii}. As a result, we were forced to compute many of the phase space integrals numerically. In most cases, we found that the numerical results were consistent with simple analytic expressions to high precision, which suggests that an analytic formula for the large $\mu$ expansion should exist. 

There are many future applications of the formalism in this work. As discussed in the introduction, all 3d $\mathcal{N}=4$ CFTs contain a 1d topological sector that can be used to compute protected OPE coefficients in terms of the $n$-body matrix model averages that we consider. For instance, in \cite{Agmon:2019imm} the protected OPE coefficients that appear in the correlators of low lying half-BPS four-point functions in $U(N)_1\times U(N)_{-1}$ ABJM theory  were written explicitly as $(n\geq4)$-body operators. If these quantities could be computed to all orders in $1/N$ using our formalism, then the result could be compared to the numerical bootstrap result of \cite{Agmon:2019imm}, as was done for the correlator of stress tensor multiplet operators in \cite{Agmon:2017xes}. One could also use our formalism to compute the integrated correlators of $n$ conserved currents in 3d $\mathcal{N}=4$ CFTs, which as discussed in the introduction are related to $n$-body operator expectation values in the CFT's matrix model. These integrated correlators could then be used to constrain the 3d $\mathcal{N}=4$ numerical bootstrap \cite{Chang:2019dzt}, as was done in a related 2d context in \cite{Lin:2015wcg}. 

The Fermi gas expansion is ideal for studying the M-theory regime where $N$ is large and another physical parameter, such as $k$ for ABJM or $N_f$ for the $N_f$ theory, is finite. In the large $N$ and large $k$ or $N_f$ regime with finite $k/N$ or $N_f/N$, which is called the 't Hooft limit, the Fermi gas results no longer apply. Instead, other methods such as topological recursion \cite{Eynard:2008we,Eynard:2004mh} have been used to compute quantities such as the sphere partition function and the expectation value of Wilson loops \cite{Drukker:2010nc}. It would nice to apply these methods to the $n$-body operators considered in this work, which would complement the Fermi gas analysis and provide a window on a larger range of physical parameters. More ambitiously, if one could compute the non-perturbative corrections in both methods, then one might able to find exact expressions for these OPE coefficients, as was previously achieved for the ABJM sphere partition function \cite{Marino:2016new,Hatsuda:2013gj,Hatsuda:2012dt,Calvo:2012du,Drukker:2010nc,Marino:2009jd,Hatsuda:2013oxa}.

\section*{Acknowledgments} 

We thank Ofer Aharony, Nathan Agmon, Silviu Pufu, Marcos Marino, Yifan Wang, Alba Grassi, Masazumi Honda, Erez Urbach, Ohad Mamroud and Barry Bradlyn for useful conversation, and Ofer Aharony for reading through the manuscript. We also thank the organizers of “Bootstrap 2019” and Perimeter Institute for Theoretical Physics for its hospitality during the course of this work. SMC is supported by the Zuckerman STEM Leadership Fellowship. This work was supported in part
by an Israel Science Foundation center for excellence grant (grant number 1989/14).

\appendix

\section{The double sine function}
\label{special}
The double sine function $s_b(x)$ (for reviews see for instance \cite{Bytsko:2006ut,Hatsuda:2016uqa}) is defined as
\beq
s_b(x)=\exp\left[-\frac{i\pi}{2}x^2-\frac{i\pi}{24}\left(b^2+b^{-2}\right)+\int_{\mathbb{R}+i0}\frac{dt}{4t}\frac{e^{-2itx}}{\sinh\left(bt\right)\sinh\left(t/b\right)}\right]\,,
\eeq
where the integration contour evades the the pole at $t= 0$ by going into the upper half–plane.
This function obeys several identities:
\begin{enumerate}
	\item $s_b^{-1}(x)=s_b(-x)$.
	\item $s_{b^{-1}}(x)=s_b(x)$.
	\item $s_b\left(\frac{ib}{2}-\sigma\right)s_b\left(\frac{ib}{2}+\sigma\right)=\frac{1}{2\cosh(\pi b\sigma)}$.
\end{enumerate}
The last identity is important when simplifying the partition function on the round sphere $b=1$.

\section{Phase space integrals}
\label{phase}

In this appendix we give details for the phase space calculations in the main text. The analytic calculations are similar to those performed in \cite{Marino:2011eh}. We first discuss the phase space integrals for the ABJM matrix model, and then for the $N_f$ matrix model.

\subsection{ABJM}
\subsubsection{Phase space integrals for A}
\label{abjmApp}
We need to calculate the phase space integral $\text{A}$ from equation \eqref{rhosimp2}, which we rewrite here for convenience:
\es{eqapp}{
	\text{A} =-\pi\partial^2_\mu \csc(\pi \partial_\mu)\int \frac{dxdp}{2\pi\hbar}\frac{x^2}{4}\left[\theta(\mu-H_W)+\sum_{r=2}^\infty\frac{\mathcal{G}_r}{r!}\delta^{(r-1)}(\mu-H_W)\right]\,.
}
We perform the calculation in an expansion in $\hbar$. The integrals which contribute order-by-order in $\hbar$ are:
\begin{enumerate}
	\item The $\frac{1}{\hbar}$ term in $\text{A}$ is given by the integral
	\beq
	\text{A}|_{\hbar^{-1}}=-\frac{\pi}{4}\partial^2_\mu \csc(\pi \partial_\mu) \int\frac{dxdp}{2\pi\hbar}x^2\theta(\mu-H_W)|_{\hbar^0}\;.
	\eeq
	\item The $\hbar$ term is given by the integral
	\beq
	\begin{split}
	\text{A}|_{\hbar}=&
	-\frac{\pi}{4} \partial^2_\mu \csc(\pi \partial_\mu)\int\frac{dxdp}{2\pi\hbar}x^2\bigg[ \theta(\mu-H_W)|_{\hbar^2} +\\ &+\frac{\mathcal{G}_2|_{\hbar^2}}{2!} \delta'(\mu-H_W)|_{\hbar^0}+ \frac{\mathcal{G}_3|_{\hbar^2}}{3!} \delta''(\mu-H_W)|_{\hbar^0}  \bigg]\;.
	\end{split}
	\eeq
	\item Finally, the $\hbar^3$ term comes from the integral 
	\beq
	\begin{split}
		\text{A}|_{\hbar^{3}}=&-\frac{\pi}{4} \partial^2_\mu \csc(\pi \partial_\mu)\int\frac{dxdp}{2\pi\hbar}x^2\bigg[ 
		\theta(\mu-H_W)|_{\hbar^4} + \frac{\mathcal{G}_2|_{\hbar^4}}{2!} \delta'(\mu-H_W)|_{\hbar^0}+\\
		&+\frac{\mathcal{G}_2|_{\hbar^2}}{2!} \delta'(\mu-H_W)|_{\hbar^2}+\frac{\mathcal{G}_3|_{\hbar^4}}{3!} \delta''(\mu-H_W)|_{\hbar^0}+\frac{\mathcal{G}_3|_{\hbar^2}}{3!} \delta''(\mu-H_W)|_{\hbar^2}+\\
		&
		+ \frac{\mathcal{G}_4|_{\hbar^4}}{4!} \delta^{(3)}(\mu-H_W)|_{\hbar^0}
		+ \frac{\mathcal{G}_5|_{\hbar^4}}{5!} \delta^{(4)}(\mu-H_W)|_{\hbar^0}
		+ \frac{\mathcal{G}_6|_{\hbar^4}}{6!} \delta^{(5)}(\mu-H_W)|_{\hbar^0} \bigg]\;.
	\end{split}
	\eeq
\end{enumerate}
The expressions for $H_W$ and $\mathcal{G}_r$ to order $\hbar^4$ are given in Appendix \ref{hbarCorrApp}.

These integrals can be done using the methods described in \cite{Marino:2011eh}. It turns out that all Wigner-Kirkwood contributions above vanish at this order, so we can set $\mathcal{G}_i=0$. Thus the only non-vanishing phase-space integral we need to compute  to order $\hbar^4$ is
\beq n=\int dx dp x^2\theta(\mu-H_W)\;. \eeq Since we only need derivatives of this integral, we will only calculate the $\mu$-dependent terms.

\begin{figure}
	\centering
	\includegraphics[width=0.4\linewidth]{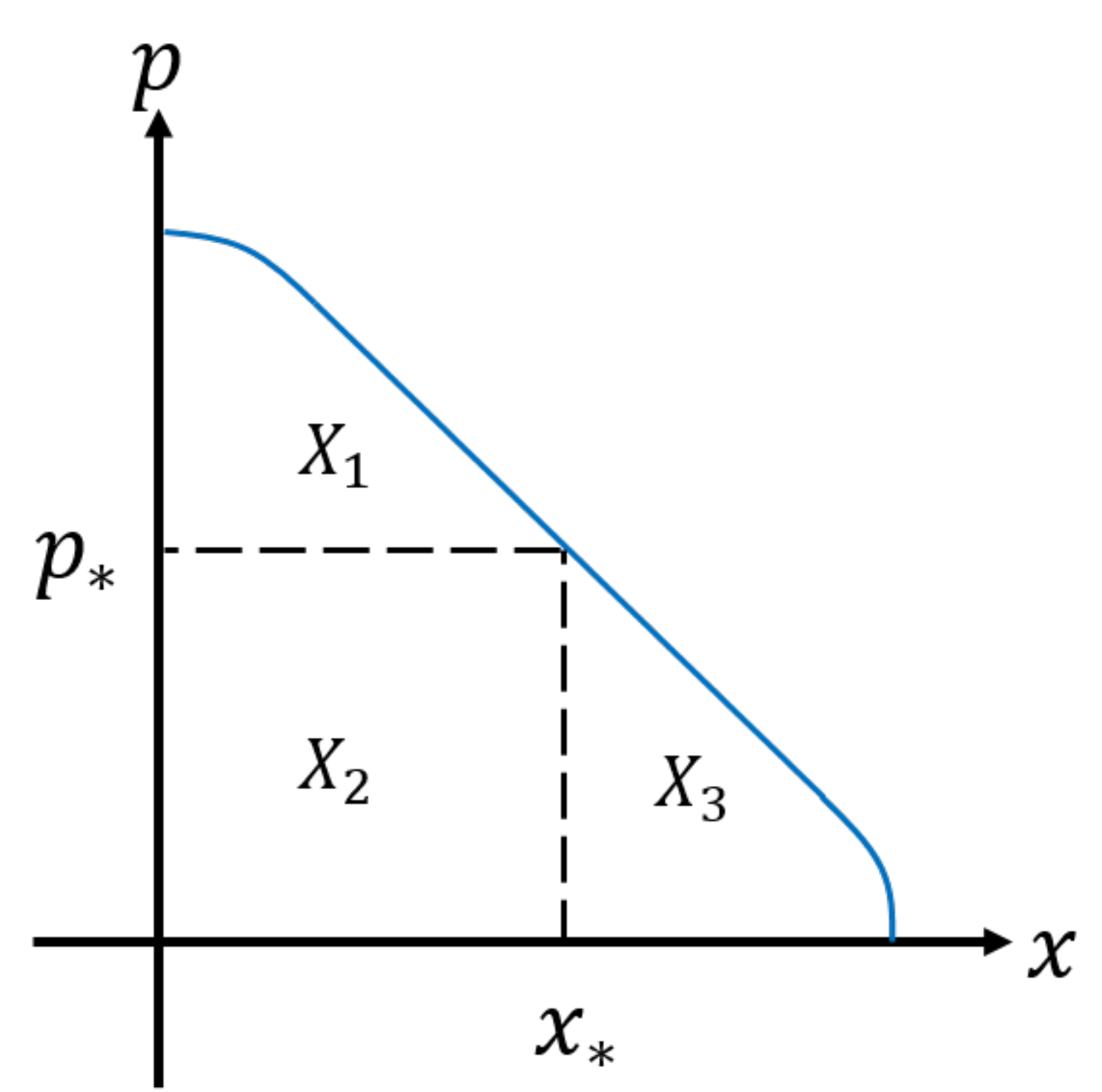}
	\caption{Integration regions for the phase-space integral.}
	\label{fig:phasespaceintegral}
\end{figure}

Following \cite{Marino:2011eh}, we can write this integral as
\beq
n=4(I_1+I_2)\;,
\eeq 
Where we have split the phase-space integral into two regions:
\beq
I_1	=\frac{1}{3}\int_{0}^{p_{*}\left(E\right)}x(\mu,p)^3dp\;,
\eeq
\beq
I_2	=\int_{0}^{x*\left(E\right)}p(\mu,x)x^2dx-\frac{1}{3}x_{*}^3p_{*}\;,
\eeq
where $p(\mu,x)$ is the solution to $H_W(x,p)=\mu$ in terms of $p$, and similarly for $x(\mu,p)$. In addition, $x_*(\mu)=p_*(\mu)=\mu$ up to exponentially small corrections. We briefly explain how this is done. First, due to symmetry under $x\to-x, p\to-p$, we can restrict ourselves to $x,p\geq 0$. We are looking for solutions to the equation 
\beq\label{eq:curve} H_W(x,p)=\mu\;,\eeq
with $H_W$ defined in Appendix \ref{hbarCorrApp}. This equation defines a curve in the space $(x,p)$. Consider the point $(x_*,p_*)$ along the curve such that $p_*=\mu$. Then from the form of $H_W$ we see that $x_*=\mu$ as well, up to exponentially small corrections in $\mu$. We plot the integration region in Figure \ref{fig:phasespaceintegral}.
The phase space integral $n$ can now be calculated as a sum of the integral over the regions $X_1,X_2,X_3$. Instead, we would like to perform the integral separately in the region where $p\leq\mu$ (which is $X_2\cup X_3$) and the region where $x\leq\mu$ (which is $X_1\cup X_2$). The integral over $X_2\cup X_3$ is precisely what we called $I_1$. As for the integral over $X_1\cup X_2$, we would like to call this $I_2$, but it is clear that we are overcounting since we are integrating over $X_2$ twice. To remove its contribution, we simply subtract the integral over $X_2$, which is simply the square integral $\int_0^{x_*} dx \int_0^{p_*} dp x^2=-\frac{1}{3}x_*^3p_*$. Subtracting this from the integral over $X_1\cup X_2$, we are left precisely with $I_2$. Thus the sum $4(I_1+I_2)$ is exactly the phase-space integral $n$.

We can find $p(\mu,x),x(\mu,p)$ using an expansion in $\hbar$ (using the expression for the Hamiltonian in Appendix \ref{hbarCorrApp}). We start by calculating 
$I_1$. Up to exponentially small corrections, we can write
\beq
x(\mu,p)=\left(2\mu-p\right)+\left(p-2T(p)\right)-\frac{\hbar^2}{48}T''(p)-\frac{7\hbar^4}{46080}T^{(4)}(p)+O(\hbar^6)\;.\eeq
We can now perform the integral $I_1$. At leading order in $\hbar$ we need to perform the following integrals:
\begin{itemize}
	\item $\frac{1}{3}\int_0^{x_*(\mu)}(2\mu-x)^3dx$, which can be calculated explicitly and produces $\frac{5\mu^4}{4}$.
	\item $3\frac{1}{3}\int_0^{x_*(\mu)}(2\mu-q)^2\left(x-2U(x)\right)dx$. This integral is more complicated, but we can extend the integration region to infinity up to exponentially small terms. Its contribution is then $-\frac{1}{3}2\pi^2\mu^2+6\mu\zeta(3)$ plus $\mu$-independent terms. 
	\item $3\frac{1}{3}\int_0^{x_*(\mu)}(2\mu-x)\left(q-2U(x)\right)^2dx$. Extending the integration limit to infinity we find that this contributes
	$\int_0^{\infty}(2\mu-x)\left(x-2U(x)\right)^2dq=
	2\zeta(3)\mu$ plus $\mu$-independent terms.
	\item $\frac{1}{3}\int_0^{x_*(\mu)}\left(x-2U(x)\right)^3dx$. Again, extending the region of integration to infinity we find that this is independent of $\mu$.
\end{itemize}
Similar considerations allow us to calculate the rest of the integrals at subleading order in $\hbar$. We find
\beq
I_1=\frac{5\mu^4}{4}+\left(-\frac{\hbar^2}{24}-\frac{2}{3}\pi^{2}\right)\mu^2+\left(8\zeta(3)+\frac{\hbar^2}{12}+\frac{\hbar^4}{5760}\right)\mu\,.
\eeq
Next we can calculate $I_2$. We have
\beq
 p(\mu,x)=(2\mu-x)+(x-2U(x))+\frac{\hbar^2}{24}U^{\prime\prime}(x)+2\frac{1}{2^{4}}\frac{\hbar^{4}}{720}U^{(4)}(x)\,,
\eeq
and we can use similar considerations to calculate $I_2$. This calculation is much simpler, since the only  $\mu$-dependence appears at leading order in $\hbar$. We find $I_2=\frac{\mu^4}{12}$ up to $\mu$-independent terms.

Summing $I_1$ and $I_2$, we find
\beq
n=\frac{16\mu^4}{3}-\left(\frac{8\pi^2}{3}+\frac{\hbar^2}{6}\right)\mu^2+\left(32\zeta(3)+\frac{\hbar^2}{3}+\frac{\hbar^{4}}{1440}\right)\mu 
\eeq
up to $\mu$-independent terms.

\subsubsection{Phase space integrals for B}

We need to calculate
\eqref{Bint}:
\es{}{
	\text{B}=\int\frac{dxdp}{2\pi\hbar}\frac{\hbar^8}{8}\sum_{r,s=0}^\infty\frac{\mathcal{G}_r\mathcal{G}_s}{r!s!}\left(\rho^{(r)}_{GC}(H_W)\partial_p^2\rho^{(s)}_{GC}(H_W)-\partial_p\rho^{(r)}_{GC}(H_W)\partial_p\rho^{(s)}_{GC}(H_W)\right)\,,
}
where $\rho_\text{GC}(H_W)=\frac{1}{e^{H_W-\mu}+1}$. The contribution to order $\hbar^4$ from this integral are:
\begin{enumerate}
	\item At order $\hbar$ the integrals that contribute are:
	\beq
	\text{B}|_{\hbar}=
	\frac{\hbar^2}{8}\int\frac{dxdp}{2\pi\hbar}
	\left(\rho_{\text{GC}}(H_W)|_{\hbar^0}\partial_p^2\rho_{\text{GC}}(H_W)|_{\hbar^0}-\partial_p\rho_{\text{GC}}(H_W)|_{\hbar^0}\partial_p\rho_{\text{GC}}(H_W)|_{\hbar^0}\right)\;.
	\eeq
	\item At order $\hbar^3$ the integrals that contribute are:
	\beq
	\begin{split}
		\text{B}|_{\hbar^3}=&
		\frac{\hbar^2}{8}\int\frac{dxdp}{2\pi\hbar}\bigg[
		\left(\partial_{p}^2\rho_{\text{GC}}(H_W)|_{\hbar^2}\right)\rho_{\text{GC}}(H_W)|_{\hbar^0}+
		\left(\partial_{p}^2\rho_{\text{GC}}(H_W)|_{\hbar^0}\right)\rho_{\text{GC}}(H_W)|_{\hbar^2}-\\
		&+\partial_{p}^2\left(\frac{\mathcal{G}_2|_{\hbar^2}}{2!}\rho^{(2)}_{GC}(H_W)|_{\hbar^0}\right)\rho_{\text{GC}}(H_W)|_{\hbar^0}
		+\left(\partial_{p}^2\rho_{\text{GC}}(H_W)|_{\hbar^0}\right)\frac{\mathcal{G}_2|_{\hbar^2}}{2!}\rho^{(2)}_{GC}(H_W)|_{\hbar^0}+\\
		&+\partial_{p}^2\left(\frac{\mathcal{G}_3|_{\hbar^2}}{3!}\rho^{(3)}_{GC}(H_W)|_{\hbar^0}\right)\rho_{\text{GC}}(H_W)|_{\hbar^0}
		+\left(\partial_{p}^2\rho_{\text{GC}}(H_W)|_{\hbar^0}\right)\frac{\mathcal{G}_3|_{\hbar^2}}{3!}\rho^{(3)}(H_W)|_{\hbar^0}-\\
		&-(\partial_p\rho_{\text{GC}}(H_W)|_{\hbar^0})^2-2\frac{\mathcal{G}_2|_{\hbar^2}}{2!}\partial_p\rho_{\text{GC}}(H_W)|_{\hbar^0}\partial_p\rho^{(2)}_{GC}(H_W)|_{\hbar^0}-\\
		&-2\frac{\mathcal{G}_3|_{\hbar^2}}{3!}\partial_p\rho_{\text{GC}}(H_W)|_{\hbar^0}\partial_p\rho^{(3)}_{GC}(H_W)|_{\hbar^0}
		\bigg]\;.
	\end{split}
	\eeq
\end{enumerate}
Unlike the calculation for $\text{A}$, in the calculation for $\text{B}$ the Wigner-Kirkwood corrections do contribute. We computed these integrals numerically for many values of $\mu$ to get the result in the main text.

\subsection{$N_f$ matrix model}
\label{nfApp}

The expectation value of the 1-body operator takes the form \eqref{intOb}:
\beq
	\frac{\langle \cO^b_1\rangle^\text{GC}}{\Xi}=-\frac{N_f}{4}\int \frac{dxdp}{2\pi\hbar}\sum_{r=0}\frac{\mathcal{G}_r}{r!}\rho^{(r)}_\text{GC}(H_W(x,p))\left(x^2+x\sinh x+\pi^2\right)\sech^2\left(\frac{x}{2}\right)\;.
\eeq
We write down explicitly the contributions at orders $\frac{1}{\hbar}$ and $\hbar$:
\begin{itemize}
	\item At order $\frac{1}{\hbar}$ we find
	\beq
	\frac{\langle \cO^b_1\rangle^\text{GC}|_{\frac{1}{\hbar}}}{\Xi}=-\frac{N_f}{4}\int \frac{dxdp}{2\pi\hbar}\rho_\text{GC}(H_W)|_{\hbar^0}\left(x^2+x\sinh x+\pi^2\right)\sech^2\left(\frac{x}{2}\right)\;.
	\eeq
	\item At order $\hbar$ we find
	\beq
	\begin{split}
	\frac{\langle \cO^b_1\rangle^\text{GC}}{\Xi}=&-\frac{N_f}{4}\int \frac{dxdp}{2\pi\hbar}
	\left(\rho_\text{GC}(H_W)|_{\hbar^2}+\frac{\mathcal{G}_2|_{\hbar^2}}{2!}\rho^{(2)}_\text{GC}(H_W)|_{\hbar^0}+\frac{\mathcal{G}_3|_{\hbar^2}}{3!}\rho^{(3)}_\text{GC}(H_W)|_{\hbar^0}\right)\\
	&\times \left(x^2+x\sinh x+\pi^2\right)\sech^2\left(\frac{x}{2}\right)\;.
\end{split}
	\eeq
\end{itemize}
The expectation value of the 2-body operator takes the form \eqref{intOb}:
\beq
	\frac{\langle \cO^b_2\rangle^\text{GC}}{\Xi}=\int \frac{d^2xd^2p}{(2\pi\hbar)^2}\sum_{r,s=0}\frac{\mathcal{G}_r\mathcal{G}_s}{r!s!}\rho^{(r)}_\text{GC}(H_W(x_1,p_1))\rho^{(s)}_\text{GC}(H_W(x_2,p_2))\Psi(x_1,p_1,x_2,p_2)\,,
\eeq
where we have defined
\beq
\begin{split}
\Psi(x_1,p_1,x_2,p_2)= & \frac12 \left({\csch}\left(x_{12}\right)\left(2x_{12}-\left(2x_{12}^2+\pi^2\right)\coth\left(x_{12}\right)+\pi^2{\csch}\left(x_{12}\right)\right)\right)+\\
&+\delta\left(x_{12}\right)\frac{\pi ^2 \left(p_{12} {\csch}\left(\frac{p_{12}}{2}\right)+1\right)}{\cosh \left(\frac{p_{12}}{2}\right)+1}\;.
\end{split}
\eeq
We write down explicitly the contributions at orders $\frac{1}{\hbar}$ and $\hbar$:
\begin{itemize}
	\item At order $\frac{1}{\hbar}$ we find
	\beq
	\frac{\langle \cO^b_2\rangle^\text{GC}}{\Xi}=\int \frac{d^2xd^2p}{(2\pi\hbar)^2}\rho_\text{GC}(H_W(x_1,p_1))|_{\hbar^0}\rho_\text{GC}(H_W(x_2,p_2))|_{\hbar^0}\Psi(x_1,p_1,x_2,p_2)\;.
	\eeq
	\item At order $\hbar$ we find
	\beq
	\begin{split}
		\frac{\langle \cO^b_2\rangle^\text{GC}}{\Xi}=&\int \frac{d^2xd^2p}{(2\pi\hbar)^2}\bigg[
		2\rho_\text{GC}(H_W(x_1,p_1))|_{\hbar^0}\rho_\text{GC}(H_W(x_2,p_2))|_{\hbar^2}\Psi(x_1,p_1,x_2,p_2)+\\
		&+		2\frac{\mathcal{G}_2|_{\hbar^2}}{2!}\rho^{(2)}_\text{GC}(H_W(x_1,p_1))|_{\hbar^0}\rho_\text{GC}(H_W(x_2,p_2))|_{\hbar^0}\\
		&+		2\frac{\mathcal{G}_3|_{\hbar^2}}{3!}\rho^{(3)}_\text{GC}(H_W(x_1,p_1))|_{\hbar^0}\rho_\text{GC}(H_W(x_2,p_2))|_{\hbar^0}
		\bigg]\Psi(x_1,p_1,x_2,p_2)\;.
	\end{split}
	\eeq
	Here we have used the symmetry under exchanging $(x_1,p_1)\leftrightarrow (x_2,p_2)$ to simplify the integrals.
\end{itemize}
The expressions for the Hamiltonian $H_W$ and the coefficients $\mathcal{G}_r$ appear in Appendix \ref{hbarCorrApp}. We computed these integrals numerically for many values of $\mu$ to get the result in the main text.

\section{$\hbar$ corrections in Fermi gas}
\label{hbarCorrApp}

We summarize the expressions required for $\hbar$ corrections in the Fermi gas expansion. As discussed in the main text, there are two sources of $\hbar$ corrections: the Wigner transform of the Hamiltonian $H_W$ and the Wigner-Kirkwood expansion.

We start by discussing the Hamiltonian $H_W$. For ABJM, defining the functions $U(x),T(p)$ as in equation \eqref{mass0H}
\beq\label{eq:U_T_app}
U(x)=\log\left(2\cosh\frac{x}{2}\right),\;\;\;\; T(p)=\log\left(2\cosh\frac{p}{2}\right)\;,
\eeq
the Hamiltonian to order $\hbar^4$ is
\beq
H_{W}=T(p)+U(x)-\frac{\hbar^2}{12}\left(T'(p)\right)^2U''(x)+\frac{\hbar^2}{24}\left(U'(x)\right)^2T''(p)+\hbar^4 H_{W}^{(2)}\;,\label{HamApp}
\eeq
where
\beq
\begin{split}
	H_{W}^{(2)}&=\frac{1}{144}T'(p)T^{(3)}(p)U^{(4)}(x)-\frac{1}{128}U'(x)U^{(3)}(x)T^{(4)}(p)-\frac{1}{240}(U'(x))^2U''(x)\left(T''(p)\right)^2+\\
	&+\frac{1}{60}\left(T'(p)\right)^2T''(p)\left(U''(x)\right)^2-\frac{1}{80}\left(U'(x)\right)^2U''(x)T'(p)T^{(3)}(p)+\frac{1}{120}\left(T'(p)\right)^2T''(p)U'(x)U^{(3)}(x)+\\
	&+\frac{7}{5760}(U'(x))^{4}T^{(4)}(p)-\frac{1}{720}\left(T'(p)\right)^{4}U^{(4)}(x)\,.
\end{split}
\eeq
This result can be applied to the $N_f$ matrix model by simply replacing $U(x)\to  N_f U(x)$.

Next we discuss the Wigner-Kirkwood expansion at order $\hbar^4$. The Wigner-Kirkwood expansion expresses the Wigner transform of any function $f(\hat H)$ in terms of the Wigner transform of $\hat H$:
\es{WKapp}{
	f(\hat H)_W=\sum_{r\geq0}\frac{f^{(r)}(H_W)}{r!}\mathcal{G}_r\,,\qquad \mathcal{G}_r=\left[\left(\hat H-H_W(x,p)\right)^r\right]_W\,.
}
The lowest Wigner-Kirkwood coefficients $\mathcal{G}_r$ are trivially
\es{GtrivialApp}{
	\mathcal{G}_0=1\,,\qquad \mathcal{G}_1=0\,,
}
while $\mathcal{G}_r$ for $r\geq2$ can be computed using the product rule \eqref{wigProd} and have an $\hbar$ expansion
\es{GnontrivialApp}{
	\mathcal{G}_r=\sum_{n\geq\left[\frac{r+2}{3}\right]}\hbar^{2n}\mathcal{G}_r^{(n)}\,,\qquad r\geq2\,.
}
The non-vanishing coefficients to order $\hbar^4$ are
\beq
\begin{split}
	\mathcal{G}_0=&1\,,\\
	\mathcal{G}_1=&0\,,\\
	\mathcal{G}_2=&\frac{1}{192} \hbar^2 \left(\hbar^2 T^{(4)}(p) U^{(4)}(x)-48 T''(p) U''(x)\right)\,,\\
	\mathcal{G}_3=&\frac{1}{192} \hbar^2 \Big[4 \hbar^2 T^{(4)}(p) U^{(3)}(x) U'(x)+9 \hbar^2 \left(T^{(4)}(p) U''(x)^2+U^{(4)}(x) T''(p)^2\right)+\\
	&4 \hbar^2 T^{(3)}(p) U^{(4)}(x) T'(p)-48 T''(p) U'(x)^2-48 T'(p)^2 U''(x)\Big]\,,\\
	\mathcal{G}_4=&\frac{1}{16} \hbar^4 \Big[T''(p)^2 \left(5 U''(x)^2+4 U^{(3)}(x) U'(x)\right)+\\
	&+2 U^{(4)}(x) T'(p)^2 T''(p)+2 U''(x) \left(T^{(4)}(p) U'(x)^2+2 T^{(3)}(p) T'(p) U''(x)\right)\Big]\,,\\
	\mathcal{G}_5=&\frac{1}{16} \hbar^4 \Big[T^{(4)}(p) U'(x)^4+U^{(4)}(x) T'(p)^4+18 T'(p)^2 T''(p) U''(x)^2+4 U^{(3)}(x) T'(p)^2 T''(p) U'(x)\\
	&2 U'(x)^2 U''(x) \left(9 T''(p)^2+2 T^{(3)}(p) T'(p)\right)\Big]\,,\\
	\mathcal{G}_6=&\frac{5}{8} \hbar^4 \left(T''(p) U'(x)^2+T'(p)^2 U''(x)\right)^2\,.
\end{split}
\eeq

\bibliographystyle{ssg}
\bibliography{fermi}

\end{document}